# Field-effect transistors of high-mobility few-layer SnSe$_2$

Chenglei Guo,[1,2] Zhen Tian,[1,2] Yanjun Xiao,[2] Qixi Mi[2] and Jiamin Xue[1,2,3,a]

[1] *Shanghai Institute of Optics and Fine Mechanics, Chinese Academy of Sciences, Shanghai 201800, China*

[2] *School of Physical Science and Technology, ShanghaiTech University, Shanghai 201210, China*

[3] *Center for Excellence in Superconducting Electronics (CENSE), Chinese Academy of Sciences, Shanghai 200050, China*

Abstract

We report the transport properties of mechanically exfoliated few-layer SnSe$_2$ flakes, whose mobility is found with four probe measurements to be ~ 85 cm$^2$V$^{-1}$s$^{-1}$ at 300 K, higher than those of the majority of few-layer transitional metal dichalcogenides (TMDs). The mobility increases strongly with decreased temperature, indicating a phonon limited transport. The conductivity of the semiconducting SnSe$_2$ shows a metallic behavior, which is explained by two competing factors involving the different temperature dependence of mobility and carrier density. The Fermi level is found to be 87 meV below the conduction band minima (CBM) at 300 K and 12 meV below the CBM at 78 K, resulting from a heavy n-type doping. Previous studies have found SnSe$_2$ field-effect transistors (FETs) to be very difficult to turn off. We find the limiting factor to be the flake thickness compared with the maximum depletion width. With fully depleted devices, we are able to achieve a current on-off ratio of ~10$^5$. These results demonstrate the great potential of SnSe$_2$ as a two dimensional (2D) semiconducting material and are helpful for our understanding of other heavily doped 2D materials.

[a] xuejm@shanghaitech.edu.cn

Two dimensional transition metal dichalcogenides have drawn great attention due to their unique mechanical, optical and electronic properties. These materials have a general chemical formula $MX_2$, where M represents a transition metal (Nb, Ta, Mo, W, Re, etc.) atom and X a chalcogen (S, Se) element. Recent studies showed their applications in high-speed, low-power FETs,[1] phototransistors,[2] logic circuits,[3] etc. While TMDs have been extensively investigated, group IVA dichalcogenides, such as $SnSe_2$, have just started to attract attention. Thick (84 nm) $SnSe_2$ flakes have been made into high drive current FETs.[4] Thinner $SnSe_2$ flakes in combination with other 2D materials, such as black phosphorus (BP)[5] and $WSe_2$,[6] have been adopted in tunneling devices, which showed pronounced negative differential resistance (with BP and $WSe_2$) or good subthreshold swing (with $WSe_2$). $SnSe_2$ flakes have also been used as high performance photodetectors.[7,8] Despite these efforts in utilizing $SnSe_2$, however, most of its basic electronic properties as a 2D material still remain to be explored.

In this work, we investigate the transport properties of few-layer $SnSe_2$ flakes exfoliated from bulk single crystals. We find their phonon limited intrinsic mobility to be ~ 85 $cm^2V^{-1}s^{-1}$ at 300 K and 233 $cm^2V^{-1}s^{-1}$ at 78 K, which is higher than those of most of the TMDs studied to date.[9] We also measure their temperature dependent carrier density, from which we carefully extract the Fermi level position to be 87 meV below the CBM at 300 K and 12 meV below the CBM at 78 K. The small activation energy combined with a strong temperature dependence of mobility results in an interesting metallic behavior. Due to its heavy n doping, FETs based on $SnSe_2$ have been found to be very difficult (if not impossible) to turn off. [4-7,10] We solve this problem with a vertical charge distribution model and identify a critical flake thickness, below which the FET can be turned off with a simple Si back gate. With this type of device, we are able to achieve a current on-off ratio of ~$10^5$ at 78 K.

The crystal structure of SnSe$_2$ belongs to the hexagonal space group P$\bar{3}$m1. It is a van der Waals material with 1T structure. Within each layer, every six Se atoms are located at the corners of an octahedron and feature an inversion symmetry with respect to the central Sn atom (Figure 1(a) inset). We grew bulk SnSe$_2$ single crystals with the chemical vapor transport method.[11] Tin powder (99.99%, Macklin) and selenium powder (99.999%, Aladdin) were mixed with a stoichiometric ratio of 1:2, without further purification. Iodine prills (99.8%, Greagent) were used as the transport agent. The mixture was sealed in a quartz tube under the pressure of $10^{-3}$ Pa and placed in a two-zone furnace, where the mixture at one end was kept at 550 ℃ and the growth end was set at 500 ℃. After three days, large (~dozens of mm$^2$) and thin (~100 μm) pieces of black SnSe$_2$ flakes with metallic luster were obtained at the cold end of the tube.

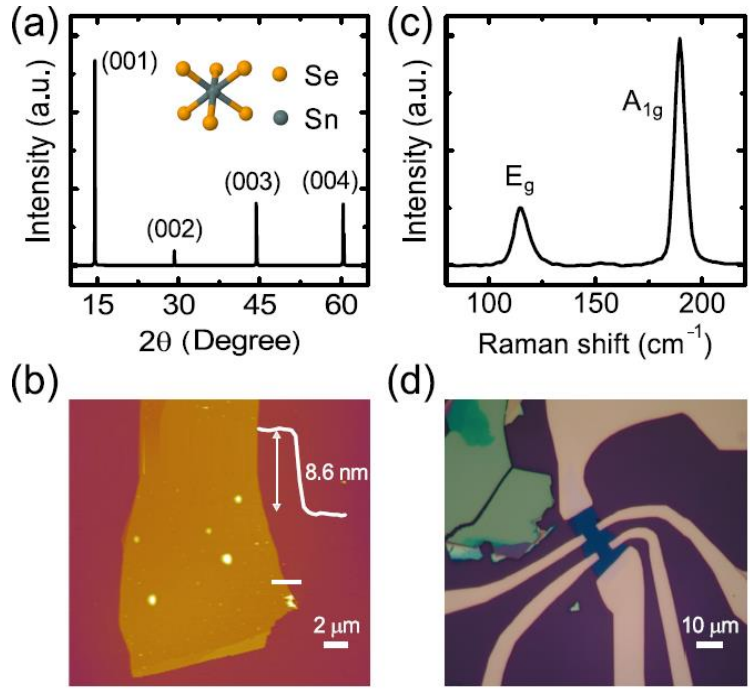

FIG. 1. (a) XRD data of an as-grown SnSe$_2$ crystal. Only peaks corresponding to {001} planes are visible due to the highly oriented van der Waals structure. (b) AFM image of an 8.6 nm thick SnSe$_2$ flake. (c)

Raman spectrum of the same sample in (b). (d) Optical microscope image of the 8.6 nm SnSe$_2$ FET device on 300 nm silicon dioxide substrate.

In order to examine the quality of the as-grown crystals, X-ray diffraction (XRD) (Bruker, D2-PHASER) was performed. XRD data (Figure 1(a)) indicates that our SnSe$_2$ crystal has highly oriented layers with an interlayer distance of 0.614 nm. The diffraction peaks match well with previous SnSe$_2$ powder XRD results.[8,12] We then used the scotch tape method[13] to exfoliate SnSe$_2$ flakes onto highly doped p-type Si substrates covered with 300 nm thermally grown SiO$_2$. Few-layer (10 nm or less) flakes were identified with optical microscope based on the interference colors of the material on the oxide substrate.[14] Then atomic force microscope (AFM) (Asylum, MFP-3D) was used to measure their thickness. One typical few layer flake of 8.6 nm is shown in Figure 1(b). Based on the interlayer distance measured by XRD, this flake consists of ~14 layers.

To examine the quality and stability of the exfoliated flakes, we use Raman spectroscopy (Thermo Fisher Scientific, DXR) to measure the fingerprint vibrational modes of the sample. Figure 1(c) depicts the Raman spectrum of the flake shown in Figure 1(b), using 532 nm laser excitation with 2 mW incident power. There are two strong peaks located at 110.2 cm$^{-1}$ and 185.4 cm$^{-1}$, corresponding to the E$_g$ (in plane) vibration mode and the A$_{1g}$ (out of plane) vibration mode, in good agreement with bulk single crystals.[15] In addition, the Raman signal of few-layer SnSe$_2$ flakes did not show any detectable degrading even when the samples were stored in ambient condition for weeks, manifesting the stability of SnSe$_2$ at its 2D limit. Previous experimental and theoretical studies have shown that SnSe$_2$ has an indirect band gap.[16-18] We attempted to measure photoluminescence (PL) on the exfoliated thin SnSe$_2$ flakes. No PL peak near its band gap[16] of 1.0 eV could be detected, which is in line with the gap being indirect.

To assess the electronic properties of SnSe$_2$ few-layer flakes, we used electron-beam lithography to fabricate SnSe$_2$ FETs. An optical image of the 8.6 nm thick SnSe$_2$ FET device is shown in Figure 1(d). 5 nm of titanium and 50 nm of gold were deposited as electrodes by electron-beam evaporation followed by lift off. All electrical measurements were performed in vacuum (~10$^{-5}$ mbar) and in dark with a Janis ST-500 probe station.

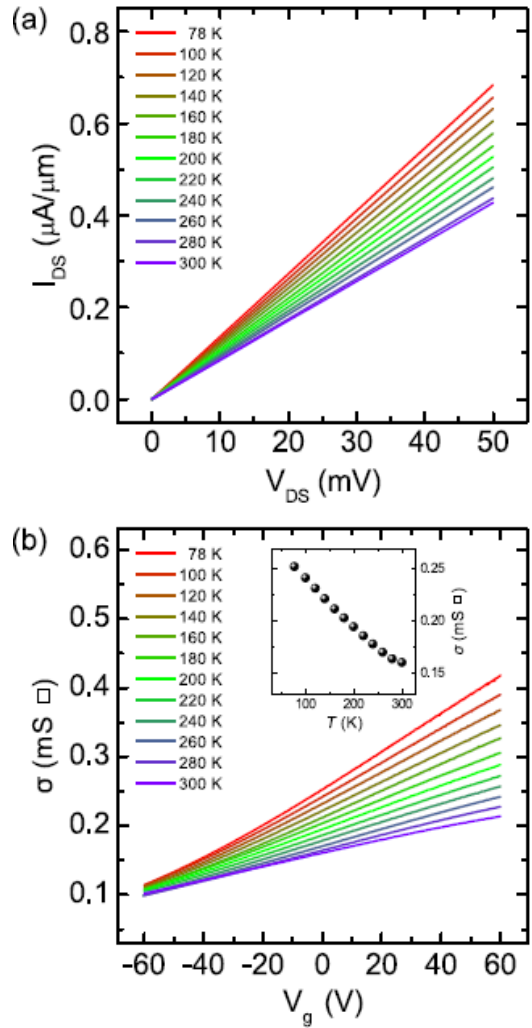

FIG. 2. (a) $I_{DS}$ vs. $V_{DS}$ of the 8.6 nm SnSe$_2$ FET device at various temperatures $T$ with the back-gate grounded. (b) Intrinsic conductivity $\sigma$ vs. back-gate voltage $V_g$ of the 8.6 nm SnSe$_2$ FET device at various

*T*. Inset: zero-gate-voltage conductivity extracted from Fig. 2(b) main panel data increases with decreasing temperature, presenting metallic behavior in the semiconducting SnSe$_2$.

Figure 2 shows electrical transport characteristics of the back-gated SnSe$_2$ FET device shown in Figure 1(d), with channel length $L$ = 6.0 μm and width $W$ = 5.7 μm. Two terminal *I-V* curves measured at different temperatures with back gate grounded is shown in Figure 2(a). The drain-source current $I_{DS}$ changes linearly with the drain-source voltage $V_{DS}$ from room temperature all the way down to 78 K, indicating the good electrical contact. With a work function of 4.3 eV,[19] titanium is expected to form an Ohmic contact with n-type SnSe$_2$, which has an electron affinity of 5.2 eV.[20] Figure 2(b) shows the intrinsic conductivity $\sigma$ versus back-gate voltage $V_g$ using four-probe measurement, with temperatures ranging from 300 K to 78 K. The device shows an n-type behavior, consistent with SnSe$_2$ being an n-type semiconductor and titanium electrodes having no Schottky barriers to its conduction band. A couple of interesting features are noticeable in the data. First, in the whole gate voltage range from - 60 to 60 V the transfer curve is almost linear at all temperatures. No turn off can be observed. This is due to the very high carrier density, which will be discussed in more details later. Second, as temperature goes down the conductivity increases. To see it more clearly, we plot the zero-gate-voltage conductivity at various temperatures in the inset of Figure 2(b). This seemingly metallic behavior is actually due to two competing factors that determine the conductivity. In the simple Drude model conductivity $\sigma = n\mu e$, where $\mu$ is the carrier mobility, $n$ the carrier density and $e$ the electron charge. For a semiconductor with electron-phonon scattering limited mobility, $\mu$ increases with decreasing temperature $T$ as $\mu \propto T^{-\gamma}$, where theoretical predicted $\gamma$ is about 1.7.[21] On the other hand, $n$ decreases with decreasing $T$ as $n \propto T^{1.5}\exp[-(E_C - E_F)/kT]$,[22] where $E_F$ is the Fermi level, $E_C$ the CBM energy and $k$ the Boltzmann constant. For a heavily doped semiconductor, $E_C - E_F$ is small, so the dependence of $\mu$ on $T$ dominates. This will result in increasing $\sigma$ with decreasing $T$, contradicting to the behavior of a normal semiconductor.

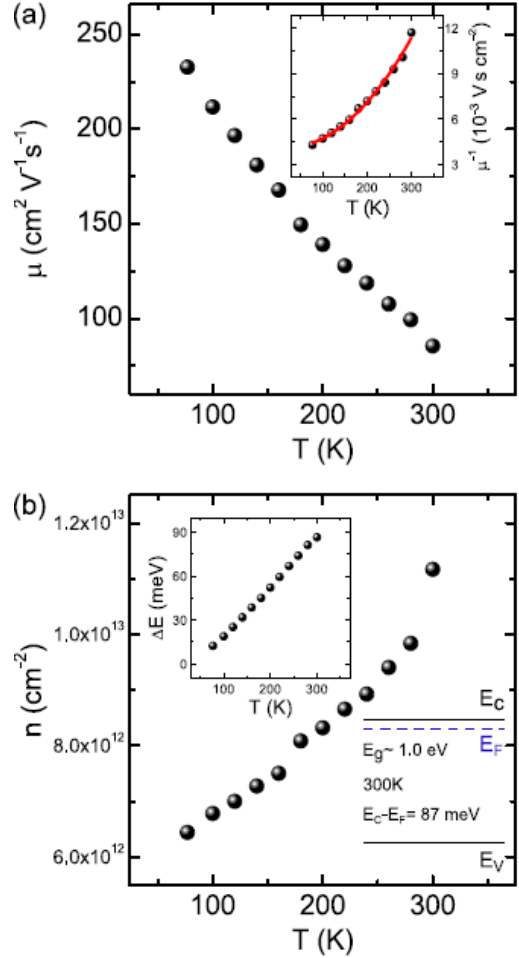

FIG. 3. (a) Extracted intrinsic field-effect mobility $\mu$ vs. $T$. Inset: fitting Fig. 3(a) main panel data with the form $\mu^{-1} = A + BT^\gamma$ yields an exponent $\gamma = 2.1$, indicating that phonon scattering limits the mobility. (b) Extracted carrier density $n$ vs. $T$. Upper inset: the energy difference between the Fermi level and the CBM $\Delta E = E_C - E_F$ at various $T$. Lower inset: band diagram of SnSe$_2$ at 300K with $\Delta E = 87$ meV.

We further quantitatively analyze this behavior by extracting the field-effect mobility $\mu$ and carrier density $n$ at different temperatures. Temperature-dependent intrinsic mobility of the 8.6 nm SnSe$_2$ FET device is shown in Figure 3(a). The mobility at room temperature is found to be 85 cm$^2$V$^{-1}$s$^{-1}$, extracted from $\mu = (d\sigma/dV_g)/C_{ox}$, where $C_{ox} = 1.15\times10^{-8}$ F/cm$^2$ is the gate oxide capacitance per unit area. The mobility compares favorably with other few-layer 2D materials, such as MoS$_2$, whose room-temperature mobility is about 50 cm$^2$V$^{-1}$s$^{-1}$ without interface engineering.[9] This shows the great potential of SnSe$_2$ as a high quality channel material for ultra-thin FETs. From analyzing the temperature dependence of $\mu$, the major scattering mechanism of electrons in SnSe$_2$ can be found. Phonon and charged impurity scattering are the two major factors that affect mobility, which have distinctly different temperature dependence. The data in Figure 3(a) shows increased $\mu$ at decreased $T$, which can be fitted with the formula $\mu^{-1} = A + BT^{\gamma}$, as shown in Figure 3(a) inset. The exponent $\gamma = 2.1$ is obtained, indicating electron-phonon scattering to be the dominant factor that limits the mobility of SnSe$_2$ devices.

We then analyze the temperature dependence of carrier density at zero gate voltage, from which the energy difference $\Delta E$ between the Fermi level and the CBM at various temperatures is obtained. The carrier density can be calculated as $n = \sigma/(\mu e)$, where both $\sigma$ and $\mu$ are zero back-gate voltage values. Figure 3(b) shows that the extracted carrier density decreases from 300 K to 78 K, which is expected for a semiconductor. On the other hand, for a true metallic material the carrier density should not change with temperature[23] (except for a minor change due to thermal expansion of the material). With the knowledge of $n(T)$ in Figure 3(b), $\Delta E$ can be derived. Considering SnSe$_2$ as nondegenerate (which is justified later), the carrier density can be expressed as[22]

$$n = 2M_C(\frac{2\pi m_{de}kT}{h^2})^{3/2}\exp(-\frac{E_C-E_F}{kT}), \quad (1)$$

where $h$ is the Plank constant, $m_{de}$ the density-of-states effective mass at the CBM (or valley) and $M_C$ the number of valleys in the conduction band within the first Brillouin zone. For SnSe$_2$ the CBMs are located at the edges of the hexagonal prismatic Brillouin zone,[17] so there are 3 degenerate valleys and $M_C = 3$. The density-of-states effective mass for SnSe$_2$ at CBM can be taken as 2.9m$_0$ (m$_0$ is the free electron mass).[24] Combining these values with Eq. (1), the energy difference $\Delta E = E_C - E_F$ is obtained, as shown in Figure 3(b) upper inset. At elevated $T$, the carriers induced by the thermal excitation slightly move the Fermi level down to promote more electrons from the donor level to the conduction band, similar with previous result for silicon.[22] In Figure 3(b) lower inset, we draw the band diagram with $\Delta E$ of 87 meV at room temperature. This value is small compared with the band gap (but still more than 2 times larger than the thermal energy, justifying the nondegenerate assumption used for Eq. (1)), indicating high doping density and shallow donor level in the SnSe$_2$ flakes. The candidates for the dopants could be intrinsic, like Sn vacancies; or extrinsic, like I impurities. Further study is needed to fully understand the doping mechanism and improve the crystal quality.

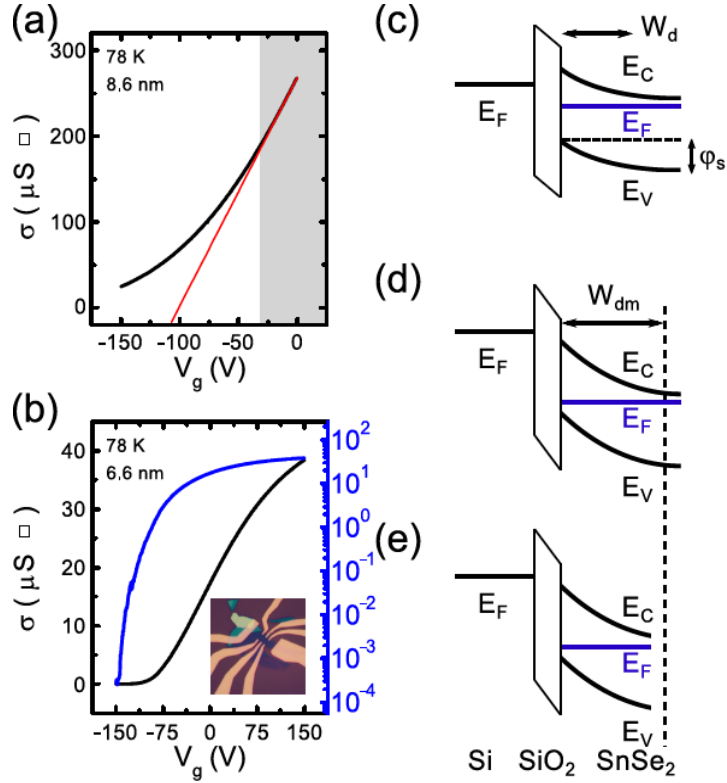

FIG. 4. (a) $\sigma$ vs. $V_g$ for the 8.6 nm thick SnSe$_2$ FET device at 78 K. Shaded area in the curve shows linear $\sigma$ vs. $V_g$ regime with small negative $V_g$. (b) $\sigma$ vs. $V_g$ for a 6.6 nm thick SnSe2 device at 78 K. (c) and (d) Band diagrams in the Si/SiO$_2$/SnSe$_2$ structure with small negative $V_g$ and large negative $V_g$ respectively. The flake is thicker than the maximum depletion width $W_{dm}$. (e) Band diagram for a fully depleted sample with thickness less than $W_{dm}$.

Due to the high doping density, a large gate voltage is needed to fully deplete the channel and turn off the device. According to the data in Figure 3(b) the carrier density $n_0$ at zero gate voltage at 78 K is $6.4 \times 10^{12}$ cm$^{-2}$. Using the simple capacitance model, to deplete the device we need a gate voltage of $V_g = -n_0 e/C_{ox} \approx -90$ V. However, even when the $V_g$ is swept to -150 V the device is still not turned off as shown in Figure 4(a). The conductivity decreases at a slower rate than the gate voltage after $V_g$ passes -30 V. This behavior has been observed by several previous reports.[4-7,10] It can be understood by considering the charge distribution in the vertical direction of the flake. Due to the finite thickness and high carrier density, we cannot treat the 8.6 nm SnSe$_2$ flake as a homogenous 2D electron gas. Using the metal-insulator-semiconductor model,[22] the gate can only modulate carrier density within the depletion width (or Debye length at very small gate voltage) near the oxide. From small to medium negative gate voltage the depletion width is growing according to $W_d = \sqrt{2\varepsilon_r \varepsilon_0 \varphi_s/(eN_D)}$,[22] where $\varepsilon_r$ is the relative permittivity of semiconductor, $\varepsilon_0$ the vacuum permittivity, $\varphi_s$ the surface potential defined in Figure 4(c) and $N_D$ the donor density. This corresponds to the linear regime in shaded area of Figure 4(a), with the band diagram shown in Figure 4(c). When the gate voltage becomes even more negative, the surface of SnSe$_2$ close to the oxide becomes inverted, as shown in Figure 4(d). In this regime the gate voltage starts to induce holes in the channel instead of further depleting the electrons. Since the contacts have a large barrier to the valence band, holes cannot contribute to $I_{DS}$. This results in the conductivity deviating from the linear dependence on gate voltage. We can calculate the maximum depletion width $W_{dm}$ based on this model. Considering that the initial flat band $E_F$ is very close to $E_C$ (Figure 3(b)) the band gap can be used for the maximum $\varphi_s$, and the carrier density at 300 K can be used as a conservative estimation for $N_D$. The relative permittivity $\varepsilon_r$ of SnSe$_2$ perpendicular to the layers is 9.97.[24] These values give $W_{dm}$ = 8.9 nm. Note that this is the upper bound for $W_{dm}$ since $N_D$ is underestimated. The 8.6 nm flake is probably thicker than $W_{dm}$, so it cannot be fully depleted no matter how large the gate voltage is, as shown in Figure 4(d).

To confirm this, thinner devices with thickness less than $W_{dm}$ were fabricated. Figure 4(b) inset shows such a device with 6.6 nm SnSe$_2$. Being thinner than $W_{dm}$, the whole channel can be depleted before the inversion layer is induced (Figure 4(e)). As expected, we can turn off the device nicely with a threshold voltage of ~ -100 V. The obtainable on-off ratio in the gate voltage range is 10$^5$, which is much higher than previously reported back-gated SnSe$_2$ devices[4-8] and much easier to operate compared with ion gel gated devices.[10] However, thinner devices generally show smaller mobility probably due to stronger scattering from the substrate.[9] $\mu$ of the device in Figure 4(b) is ~20 cm$^2$V$^{-1}$s$^{-1}$. With advanced interface engineering to reduce scattering and the high mobility potential of SnSe$_2$ demonstrated in the thicker flake, we expect to obtain devices with both good mobility and low threshold voltage in the future.

In summary, we fabricate n-type FETs with heavily doped few-layer SnSe$_2$, which shows the highest on-off ratio up to ~10$^5$ at 78K. Using four-probe measurement, intrinsic field-effect mobility has been obtained, with the highest value up to 85 cm$^2$V$^{-1}$s$^{-1}$ at room temperature. Due to the small energy gap between CBM and $E_F$, a seemingly metallic behavior occurs in semiconducting SnSe$_2$. The mobility of SnSe$_2$ FET increases significantly with lowering temperature, suggesting phonon scattering as the main factor limiting its mobility. We want to point out that even with such a high doping level, few-layer SnSe$_2$ still shows good mobility. Currently we are working on growing higher quality crystals with less dopant concentration. With this improvement, we expect few- to mono-layer SnSe$_2$ FETs to fully demonstrate its great potential as a high-mobility material for applications in nanoelectronic devices.

We thank Prof. Xufeng Kou of ShanghaiTech University for insightful comments. This work was supported by National Natural Science Foundation of China (Grant No. 11504234), Science and Technology Commission of Shanghai Municipality (Grant Nos. 15QA1403200 and 14PJ1406600) and ShanghaiTech University. The nanofabrication facility was supported by the Strategic Priority Research Program (B) of the Chinese Academy of Sciences (Grant No. XDB04030000).

13